\documentstyle[sprocl]{article}

\def\ra{\rightarrow}

\def\be{\begin{equation}}
\def\ee{\end{equation}}
\def\bea{\begin{eqnarray}}
\def\eea{\end{eqnarray}}

\def\sqr#1#2{{\vcenter{\hrule height.#2pt
   \hbox{\vrule width.#2pt height#1pt \kern#1pt
    \vrule width.#2pt}
    \hrule height.#2pt}}}

\def\meio { \frac{1}{2} }
\def\Lie { {\cal L}_{\xi} }  
\def\gmn { g_{\mu \nu} }

\def\tq { \tilde{q} }

\def\Gmn { G_{\mu \nu} }

\def\Tmn { T_{\mu \nu} }

\def\tx { \tilde {x} }
\def\eps { \epsilon }
\def\taumn { \tau_{\mu \nu} }

\begin{document}

\title{ THE BACK REACTION OF GRAVITATIONAL PERTURBATIONS
\footnote{ {\bf BROWN-HET-1075}, to appear in proceedings of the
18$^{\rm th}$ Texas Symposium on Relativistic Astrophysics, Chicago, 1996}
}

\author{ L.R. ABRAMO, R.H. BRANDENBERGER}

\address{Physics Department, Brown University, \\ 
Providence, RI 02912, USA}

\author{ V.M. MUKHANOV }

\address{Institut f\"ur Theoretische Physik, ETH Z\"urich, \\
CH-8093 Z\"{u}rich, Switzerland }

\maketitle\abstracts{
The back reaction of gravitational perturbations in a homogeneous background
is determined by an effective energy-momentum tensor
quadratic in the perturbations. We show that this nonlinear feedback 
effect is important in the case of long wavelength scalar perturbations in 
inflationary universe models. We 
also show how to solve an old problem concerning the gauge dependence of 
the effective energy-momentum tensor of perturbations.
}

\section{Back Reaction in Chaotic Inflation}

It is a well known fact that gravitational waves carry energy
and momentum, and as such are themselves a source of curvature
for spacetime. High-frequency gravity waves, in particular,
have an equation of state of a radiation fluid ($p = \rho /3$) and 
this can be used to constrain their amplitude during BBN
by taking into account the effect they have on the expansion and cooling 
rates of the universe.

Back reaction can also be quite important in the inflationary models of the
Universe evolution. In the chaotic inflation model with a massive scalar 
\linebreak
field \cite{Chao}, for example, it is usually supposed that once the
inflaton $\varphi$ drops below the self-reproduction scale 
$\varphi_{sr} \sim m^{-1/2}$ (in Plack units), the dynamics of the 
homogeneous FRW background proceeds classically with no influence
from metric perturbations created henceforth. 
Our calculations show, however, that the effect of back reaction 
may become crucial midway through the period of slow-roll
inflation \cite{Us1,Us2}. 

To see that, consider the effective energy-momentum tensor (EEMT) 
obtained after averaging quadratic terms of the perturbative 
expansion of Einstein's Equations about the FRW background,

\be
\label{EEMT}
\taumn = \frac{1}{16 \pi} 
	\langle 
			\left( 
				\Gmn - 8 \pi \Tmn 
			\right)_{,ab} 
		\delta q^a \delta q^b 
	\rangle \,\, ,
\ee
where $\langle \dots \rangle$ denotes spatial averaging and 
$\delta q^a \equiv \{ \delta \gmn , \delta \varphi \}$ are
the perturbations to the metric and matter fields, for which 
$\langle \delta q^a \rangle =0$ but 
$\langle (\delta q)^2 \rangle \neq 0$. This energy-momentum tensor, of
second order
in the perturbations, serves as an effective source in the RHS of
Einstein's equations.

The energy density in long wavelength 
scalar (density) perturbations ({\it i.e.}, the 0-0 component of the
EEMT) is proportional to 
$\langle \delta \varphi^2 \rangle$. Making use
of the known spectrum of first order perturbations, it can be shown that
$\tau_{00}$ becomes comparable to the 
energy density of the background before the end of inflation if the intial
value of the scalar field was bigger than $\varphi_0 \sim m^{-1/3}$.

\section{Gauge Dependence of the EEMT}

Once it is established that back reaction can be relevant and that
the EEMT is the proper tool for handling it, we must deal with a
problem inherent to that tensor: namely, its gauge dependence.

On a given manifold, a coordinate transformation
$x \ra \tx = x + \xi$ ($\xi$ small and $\langle \xi \rangle=0$)
induces a gauge transformation on the tensors of that manifold
which is expressed in terms of the Lie derivative:
$q \ra \tq = q - \Lie q$. In particular, the matter and metric field
perturbations are transformed by the same law, which perturbatively
reads

\be
\label{tr:linear}
\delta {\tq}^a = \delta q^a - [ \Lie q_0 ]^a \, \, .
\ee
Since the EEMT is a function of the explicit perturbations on the
fixed background, it is clear by Eq. \ref{EEMT} that it will
change accordingly, $\tilde{\tau}_{\mu\nu} \neq \taumn$.

The question now becomes, how to calculate back reaction and be
sure that it does not include spurious ``gauge" effects? In order
to answer this we must first clarify the origin of the mystery,
{\it i.e.} why does the EEMT change while the background
apparently does not?

\section{Finite Gauge Transformations}

Back reaction is a second order effect, and therefore any 
terms of like order should be accounted for. In particular, it must
be recognized that a coordinate transformation in fact induces a
gauge transformation which involve terms of all orders in perturbation
theory. Rather than the simple law Eq. \ref{tr:linear}, valid only
up to first order, tensor fields are transformed by the Lie operator
found upon exponentiation of the Lie derivative,

\be
\label{Lie:op}
\tq = e^{-\Lie} q \, \, .
\ee
Now it becomes clear that, although to first order the background
variables do not change under a gauge transformation, to second
order they do:

\be
\label{tr:sec}
{\tq}_0^a = q_0^a + \langle \meio \Lie \Lie q_0^a - 
		    \Lie \delta q^a \rangle + {\cal O}(\eps^3) \, \, ,
\ee
where all first order terms from Eq. \ref{Lie:op} vanish by virtue of
the spatial average.

We conclude then that, although under a coordinate transformation the EEMT
change, so does the background to which it is referred. In this case we
would write Einstein's equations in another frame as

\be
\label{Pi}
\langle \tilde{\Pi} \rangle = \langle e^{-\Lie} \Pi \rangle = 0
\quad \ra \quad 
\tilde{\Pi}_0 = -\meio \langle \Pi{,ab} \delta \tq^a \tq^b \rangle
\ee
where tensor indices have been suppressed for simplicity.

Above, we have shown that the back reaction equation \ref{Pi} for
cosmological perturbations is covariant, that is, takes the same form
in any coordinate system. An obvious question is whether a gauge invariant
formulation exists. The answer is yes, and the construction is in fact
analogous to one that leads to the gauge invariant theory of linear
perturbations \cite{Us1,Us2}.

\section*{Acknowledgments} 
L.R.A. is supported by CNPq (Research Council of Brazil); L.R.A. and R.B.
are partially financed by U.S. DOE, contract DE-FG0291ER-40688, Task A;
V.M. thanks the SNF and the Tomalla foundation for financial support; the
authors also acknowledge support by NSF collaborative research award
NSF-INT-9312335.

\end{document}